# A high-temperature superconducting weak-link defined by ferroelectric field-effect

L. Bégon-Lours, V. Rouco, A. Sander, J. Trastoy, R. Bernard♣, E. Jacquet, K. Bouzehouane, S. Fusil, V. Garcia, A. Barthélemy, M. Bibes, J. Santamaría* and J.E. Villegas♦

*Unité Mixte de Physique, CNRS, Thales, Univ. Paris-Sud, Université Paris Saclay, 91767 Palaiseau, France*

**In all-oxide ferroelectric (FE) / superconductor (S) bilayers, due to the low carrier concentration of oxides compared to transition metals, the FE interfacial polarization charges induce an accumulation (or depletion) of charge carriers in the S. This leads either to an enhancement or a depression of its critical temperature depending on FE polarization direction. Here we exploit this effect at a local scale to define planar weak-links in high-temperature superconducting wires. This is realized in $BiFeO_3$ (FE)/$YBa_2Cu_3O_{7-x}$ (S) bilayers in which the remnant FE domain structure is "written" at will by locally applying voltage pulses with a conductive-tip atomic force microscope. In this fashion, the FE domain pattern defines a spatial modulation of superconductivity. This allows us to "write" a device whose electrical transport shows different temperature regimes and magnetic field matching effects that are characteristic of Josephson coupled weak-links. This illustrates the potential of the ferroelectric approach for the realization of high-temperature superconducting devices.**

♣present address : Fonctions Optiques pour les Technologies de l'Information, UMR 6082, CNRS, INSA de Rennes, Rennes, France
*permanent address : GFMC, Dpto. Fisica de Materiales, Facultad CC. Físicas, Universidad Complutense de Madrid, Madrid, Spain
♦ javier.villegas@cnrs-thales.fr

The Josephson effect [1,2] results from the coupling of two superconductors (S) across an insulator (tunnel S/I/S junction) [3] or a weak-link [4] (e.g. a normal-metal in a proximity S/N/S junction), and constitutes the basis of a number of superconductor applications such as quantum interference devices for magnetometry, digital electronics, signal processing, or medical imaging among others [5].

Josephson applications were early realized using low critical temperature ($T_C$) superconductors. However, the advent of high $T_C$ superconductivity prompted efforts to create Josephson devices using high-$T_C$ cuprates [6] capable of operating above liquid-nitrogen temperature. A versatile approach, developed during the last decade, exploits a nanoscale spatial modulation of $T_C$ in a superconducting film [7,8]. This is accomplished by controllably introducing structural disorder (mostly oxygen interstitials and vacancies) on a local scale, via masked ion irradiation or focused ion beams. That type of disorder locally depresses $T_C$, which allows "patterning" the superconducting properties of the film [9,10]. Both S/N/S weak-links [7,8] and S/I/S tunnel junctions [11] can be defined in this way. Besides yielding very reproducible Josephson characteristics [12], the key advantage of this approach is that the planar geometry allows for very large junction arrays [13], thus opening the door to a variety of superconductor electronics applications [14].

In this paper we investigate a different approach to define planar weak-links in cuprate superconducting films. The idea is to create a nanoscale spatial modulation of $T_C$ by electrostatic doping [15] through ferroelectric field-effect [16–18]. This is achieved in ferroelectric (FE)/superconductor (S) bilayers in which we combine a FE ($BiFeO_3$) whose polarization has a strong out-of-plane component [19] and a high-$T_C$ superconductor ($YBa_2Cu_3O_{7-x}$). In this FE/S structure, if the FE polarization points away from the interface, holes are accumulated in the S to screen the negative polarization charges, which enhances $T_C$ [red curve in Fig. 1(a)]. Conversely, if the FE points towards the interface, a hole depletion is induced, which leads to a depressed $T_C$ [blue curve in Fig. 1(a)]. Switching between those two remnant states is achieved by momentarily applying a gate voltage to reverse the FE polarization [16–18]. Interestingly, this type of superconductivity modulation can be produced at a local scale, as sketched in Fig 1 (b), by artificially "writing" a FE domain pattern locally applying voltages with the Conductive Tip of an Atomic Force Microscope (CT-AFM) [20,21]. One of the prospects consists in "field-effect patterning" the superconducting condensate in a film to define circuits (such as Josephson arrays) that could be reversibly reconfigured via changes in the domain pattern [20,21]. With this perspective, here we explore a circuit defined by the FE domain structure shown in Fig. 1(b). The spatial modulation of the critical temperature created by this domain structure which results in two superconductors connected by a superconducting region with a lower critical temperature $T_C$' (a S/S'/S junction). Using transport measurements we find that, within a given temperature range, the device behaves as a Josephson weak-link and shows a non-monotonic magneto-resistance related to flux quantization effects across S'. These results underscore the potential of the ferroelectric approach to tailor high-temperature superconducting circuits.

$c$-axis $YBa_2Cu_3O_{7-x}$/$BiFeO_3$ (YBCO/BFO) were grown using pulsed laser deposition on [001] $SrTiO_3$ substrates that were previously treated (chemical etching in HF followed by annealing at 1000°C and in $O_2$) to obtain an atomically flat surface. Prior to YBCO deposition – either 3 or 4 unit cells (u.c.) for the samples studied here – a 3 u.c. buffer layer of $PrBa_2Cu_3O_{7-x}$ (PBCO, a wide-gap semiconductor isostructural to YBCO) was grown, which is known to enhance the superconducting properties of ultrathin YBCO. Note that ultrathin YBCO is required since the superconductor thickness must be comparable to the Thomas-Fermi screening length (~nm) [15] to maximize field-effect doping. Both PBCO and YBCO were grown at 690°C in 0.35 mbars of pure $O_2$. Subsequently, a ~30 nm thick BFO layer (with 5% substitution of Fe by Mn to reduce leakage currents [22]) was grown at 570°C. The thickness of the different layers was adjusted from calibrated deposition rates. Further details on the growth and structural characterization of the films can be found elsewhere [19]. After growth, a multi-probe bridge [10 μm wide with voltage probes 40 μm afar, see Fig. 1 (d)] for electrical measurements was patterned using photolithography and ion beam etching. Contact pads to wire-bond YBCO through the BFO layer consist of a series of micrometric "trenches" etched across BFO and filled with evaporated Au. The ferroelectric domain structure was written by using a CT-AFM and imaged via piezoresponse force microscopy (PFM) as reported elsewhere [21]. We verified that the PFM images, and in particular the size of the written ferroelectric domains, were stable over a time-scale of a few weeks [see. Fig. 1 (c)]. After setting the desired ferroelectric state, samples were introduced in a closed-cycle refrigerator equipped with a 9 kOe electromagnet and a rotatable sample holder to perform magneto-transport measurements.

Fig 1 (a) shows resistance versus temperature for a YBCO/BFO bilayer (YBCO thickness is 3 u.c.), in three different ferroelectric states: the blue and red curves respectively correspond to the cases in which the area in between the voltage probes is homogeneously polarized "up" and "down". The black curve corresponds to a measurement after the S/S'/S structure shown in Fig. 1 (b) and (c) has been written. Note that the zero-resistance critical temperature for the S/S'/S device lies in between those for the "up" and "down" FE states (blue/red), which implies that the S' region is proximity-coupled to the S regions.

Figure 2 shows V(I) characteristics of the S/S'/S device in the 3 u.c. YBCO sample in zero applied magnetic field for a set of temperatures. Above ~34 K, we observe a nearly Ohmic tail at low currents and non-linear behavior at higher currents. For lower temperatures, the Ohmic tail is no longer visible, and yields to a steep V(I), with (seemingly) vanishing resistance in the low-current limit. Qualitatively, this is reminiscent of the behavior observed in YBCO thin films as they undergo a $2^{nd}$ order vortex glass transition [23]. However, the data depicted in Fig. 2 show significant quantitative differences from the theoretical expectations for that scenario. For instance, the crossover current $I_{nl}$ above which nonlinear behavior is observed increases with decreasing temperature – this trend is marked with a dashed line in the figure. This is opposite to the theoretically expected behavior in the case of a vortex glass transition, where $I_{nl}$ should decrease with decreasing temperature [24–26]. As we discuss further below, the different regimes observed in the V-I characteristics, and the disagreement of the observed behavior with that of single YBCO films, are connected to the weak-link behavior of the S/S'/S junction.

Magnetic field effects in the S/S'/S junction were investigated by measuring sets of V(I) characteristics as a function of the applied magnetic field H, for 31 K<T<40 K. H is applied perpendicular to the current, and nearly parallel to the YBCO *ab* planes (the misalignment is 5 ± 1 deg; the magneto-transport dependence on the field direction will be discussed below). Fig. 3 displays V(I,H) for several temperatures (see legend) in color contour plots. Interpolated I(H) for V~20 µV and V~40 µV are drawn in white color, as a guide to the eye. For all the temperatures, I(H) at constant voltage globally decreases with increasing magnetic field, yielding a nearly linear background. However, a non-monotonic modulation is superposed to that background for the measurements shown in Figs. 3 (b) and (c), which respectively correspond to 32 and 34 K. For these temperatures, I(H) show kinks (inflexion points) around |H| ~ 4 kOe (white arrows). These features are most pronounced around 34K-32K, and disappear for lower and higher temperatures. Note that the 34K-32K temperature range corresponds to the transition between V(I) showing an Ohmic tail and V(I) showing vanishing resistance in Fig. 2.

To further investigate the magnetic field modulation of dissipation and allow comparison between different samples and measurement conditions, we performed magneto-resistance measurements as a function of temperature [Fig. 4 (a)]. The resistance is calculated as R=V/I from the voltage measured at constant current. Thus, the kinks seen in I(H) (Fig. 3) appear inverted in R(H) (marked with arrows in Fig. 4). One can see that the most pronounced effects appear in the temperature range ~31K to ~34 K.

In order to investigate anisotropic effects, we performed magneto-resistance measurements as a function of the angle θ between H and the YBCO *ab* planes, keeping the current perpendicular to H. Some of those measurements are shown in Fig 4 (b). Note that the kinks appear only in the angular range 0º<θ<30º. As the magnetic field is rotated out-of-plane (i.e. on increasing θ), those features rapidly shift to lower fields and become weaker. Note also that the background magneto-resistance becomes much stronger with increasing θ, as expected from the very anisotropic nature of ultrathin YBCO [27]. The inset of Fig. 4 (b) displays the characteristic fields at which the kinks appear, as a function of θ. Each kink is characterized by two fields, which correspond to the two inflexion points respectively marked with a downwards arrow (↓) and an upwards arrowhead (∧) on the curves in the main panel. The trend observed in the inset of Fig. 4 (b) suggests that the magneto-resistance kinks are not observed for θ = 0º because the characteristic fields grow out of the experimentally accessible window. Assuming that characteristic fields are related to a flux matching phenomena, we sought to decrease it by increasing the sample section in parallel field. With this aim, we performed experiments in a YBCO/BFO bilayer with thicker YBCO – nominally 4 u.c. – in which a S/S'/S junction was written similar as in the 3 u.c. sample. For the 4 u.c. sample, see Fig. 4 (c), the magneto-resistance measurements in strictly parallel field (θ = 0º) did show matching effects within the experimentally accessible field range. In particular resistance maxima (marked with downward arrows) are observed for |H|~5 kOe. Just as for the junction with 3 u.c.-thick YBCO, the non-monotonic R(H) is observed only within a very narrow (a

few K) range of temperatures. The characteristic matching field can be related to the section of the S' region via flux quantization arguments. In particular, using the thickness of the film $t \sim 4.4 \pm 1.1$ nm and the length of the S' barrier, $l = 1 \pm 0.1$ µm, we estimate a characteristic field B = $\phi_0/tl$ = 4.7 ± 1.3 kGauss (with $\phi_0$=2.07 $10^{-15}$ Wb), in good agreement with the experimental one H ~ 5 kOe. If we perform an estimate for the 3 u.c. sample, a characteristic field of the order of B = $\phi_0/tl$ = 6 ± 2 kGauss is obtained. This suggests that for this sample we cannot clearly see the magneto-resistance kink in strictly parallel field because it extends out of the experimental window.

The observed behaviors can be understood if we consider that, as temperature is decreased, the S/S'/S junction effectively behaves as i) decoupled S/N/S junction, ii) a proximity coupled S/N/S junction or iii) a continuous superconducting film. If we take as an example the measurements in the 3 u.c. thick YBCO, the different regimes can be described as follows:

i) At high temperatures (down to ~38 K in the example), the S/S'/S device behaves as decoupled S/N/S junction. The measured resistance corresponds to the addition of the resistances of S and S' in series. In this temperature range, S has not reached the zero-resistance state and, since the ratio between the lengths of S and S' is ~40 µm / 1 µm, the device resistance is essentially dominated by S. Thus, the resistance decreases with decreasing temperature and increases with increasing magnetic field. In this regime, due to the absence of coupling across S', I(H) and R(H) show no field matching effects.

ii) At intermediate temperatures (~32K<T<~38K in the example) the S electrodes approach the zero-resistance state and the system eventually starts behaving as a coupled S/N/S junction. In this regime, I(H) and R(H) show kinks related to flux quantization effects across S', which suggest Josephson coupling. In this scenario, we expect that the contribution of S and S' to the device resistance depends on the injected current and applied field. At low currents, i.e. below the critical current in S, the resistance is governed by the proximitized S' barrier. However, as the current and/or the magnetic field are increased, the contribution of the S electrodes increases. This current and field dependent weighting of the contribution in S and S' to the measured resistance explains i) the unusual temperature dependence of the crossover current $I_{nl}$ (the nearly Ohmic tail corresponds to the response of S' above its critical current) and ii) the weakening of the field-matching effects with increasing current, which occurs as this exceeds the critical current in S thereby increasing its contribution to the measured resistance. In consistence with this scenario, the field-matching effects become stronger at temperatures below 34 K [see Fig. 3 and 4(a)]. This corresponds to the temperature range in which V(I) becomes non-linear within the whole experimental window (see Fig. 2), due to the increase of the critical current in the proximitized S' barrier, thus leading to stronger magnetic-field modulation effects.

iii) At the lowest temperatures (below 32 K in the example) the intrinsic critical temperature of the S' barrier is reached and the device does not behave as a coupled S/N/S junction anymore, but as a continuous S/S'/S superconducting film. Thus, R(H) and I(H) do not show field matching effects related to Josephson coupling anymore.

In the above scenario, the rapid decrease of the matching field with increasing θ can be qualitatively understood considering that the effective section of the junction strongly depends on out-of-plane component of the applied magnetic field: while for the in-plane field component the junction area is given by the superconductor thickness times the length of S' (as discussed previously), for the perpendicular component the area is defined by the length of S' times the width of the transport bridge (10 μm). Thus, the effective junction area sharply increases as magnetic field is tilted out of the in-plane direction, which yields the strong decrease of matching fields observed experimentally (see Figure 4).

In summary, we have electrostatically defined proximity coupled S/S'/S weak-links in high-temperature superconducting wires by tailoring the domain structure of a ferroelectric overlayer. In particular, the local direction of the ferroelectric polarization was set so that the induced charge accumulation yields a local depression of superconductivity over a micronsize section of the wire. The field-effect patterned devices show field-matching effects in the magneto-resistance, which are the signature of flux quantization effects characteristic of a proximity coupled weak-link. The micrometric width of the S' weak-link sets a narrow temperature range over which the Josephson coupling between the S banks may occur. In particular, the temperature range limited on the upper side by the required divergence of the of coherence length of Andreev pairs in S', and on the lower side by the intrinsic critical temperature of S'. The present experiments illustrate the potential of the ferroelectric field effect approach for patterning high-temperature superconducting oxide devices.

**Acknowledgements**

Work supported by the ERC grants Nº 64710 and N° 267579 and French ANR grant ANR-15-CE24-0008-01. L. B.-L. acknowledges a doctoral fellowship from DIM "OXYMORE" (Region Ile-de-France). J. S. thanks the CNRS "Institut de Physique" and Université Paris-Saclay (Programme Jean d'Alembert) for supporting his stay at Unité Mixte de Physique CNRS/Thales. J.T. thanks support from Fundación Barrié (Galicia, Spain). We thank J. Lesueur, N. Bergeal and S. Bergeret for useful discussions.

[1]     B. D. Josephson, Phys. Lett. 1, 251 (1962).

[2]     B. D. Josephson, Adv. Phys. 14, 419 (1965).

[3]     P. W. Anderson and J. M. Rowell, Phys. Rev. Lett. 10, 230 (1963).

[4]     K. K. Likharev, Rev. Mod. Phys. 51, 101 (1979).

[5]     H. Rogalla and P. H. Kes, editors , *100 Years of Superconductivity* (Taylor & Francis, 2011).

[6]     D. Koelle, F. Ludwig, P. Bundesanstalt, S. Cryosensors, D.- Berlin, E. Dantsker, and J. Clarke, Rev. Mod. Phys. 71, 631 (1999).

[7]     K. Chen, S. A. Cybart, and R. C. Dynes, Appl. Phys. Lett. 85, 2863 (2004).

[8]     N. Bergeal, X. Grison, J. Lesueur, G. Faini, M. Aprili, and J. P. Contour, Appl. Phys. Lett. 87, 102502 (2005).

[9]     J. Trastoy, M. Malnou, C. Ulysse, R. Bernard, N. Bergeal, G. Faini, J. Lesueur, J. Briatico, and J. E. Villegas, Nat. Nanotechnol. 9, 710 (2014).

[10]    J. Trastoy, Phys. Rev. Appl. 4, (2015).

[11]    S. A. Cybart, E. Y. Cho, T. J. Wong, B. H. Wehlin, M. K. Ma, C. Huynh, and R. C. Dynes, Nat. Nanotechnol. 10, 598 (2015).

[12]    M. Sirena, X. Fabrèges, N. Bergeal, J. Lesueur, G. Faini, R. Bernard, and J. Briatico, Appl. Phys. Lett. 91, 262508 (2007).

[13]    S. A. Cybart, S. M. Anton, S. M. Wu, J. Clarke, and R. C. Dynes, Nano Lett. 9, 3581 (2009).

[14]    S. Ouanani, J. Kermorvant, C. Ulysse, M. Malnou, Y. Lemaître, B. Marcilhac, C.

Feuillet-Palma, N. Bergeal, D. Crété, J. Lesueur, and 1, Supercond. Sci. Technol. 94002 (2016).

[15]   J. Mannhart, Supercond. Sci. Technol. 9, 49 (1996).

[16]   C. H. Ahn, S. Gariglio, P. Paruch, T. Tybell, L. Antognazza, and J. M. Triscone, Science (80-. ). 284, 1152 (1999).

[17]   C. H. Ahn, J.-M. J.-M. M. Triscone, and J. Mannhart, Nature 424, 1015 (2003).

[18]   C. H. Ahn, A. Bhattacharya, M. Di Ventra, J. N. Eckstein, C. Daniel Frisbie, M. E. Gershenson, A. M. Goldman, I. H. Inoue, J. Mannhart, A. J. Millis, A. F. Morpurgo, D. Natelson, and J. M. Triscone, Rev. Mod. Phys. 78, 1185 (2006).

[19]   A. Crassous, R. Bernard, S. Fusil, K. Bouzehouane, J. Briatico, M. Bibes, A. Barthélémy, and J. E. Villegas, J. Appl. Phys. 113, 24910 (2013).

[20]   K. S. Takahashi, M. Gabay, D. Jaccard, K. Shibuya, T. Ohnishi, M. Lippmaa, and J. M. Triscone, Nature 441, 195 (2006).

[21]   A. Crassous, R. Bernard, S. Fusil, K. Bouzehouane, D. Le Bourdais, S. Enouz-Vedrenne, J. Briatico, M. Bibes, A. Barthélémy, and J. E. Villegas, Phys. Rev. Lett. 107, 247002 (2011).

[22]   X. H. Zhu, H. Bea, M. Bibes, S. Fusil, K. Bouzehouane, E. Jacquet, A. Barthelemy, D. Lebeugle, M. Viret, and D. Colson, Appl. Phys. Lett. 93, 82902 (2008).

[23]   J. Trastoy, V. Rouco, C. Ulysse, R. Bernard, A. Palau, T. Puig, G. Faini, J. Lesueur, J. Briatico, and J. E. Villegas, New J. Phys. 15, 103022 (2013).

[24]   M. P. Fisher, Phys. Rev. Lett. 62, 1415 (1989).

[25]   D. S. Fisher, M. P. A. Fisher, and D. A. Huse, Phys. Rev. B 43, 130 (1991).

[26]   R. H. Koch, V. Foglietti, W. J. Gallagher, G. Koren, A. Gupta, and M. P. A. Fisher, Phys. Rev. Lett. 63, 1511 (1989).

[27]   J. E. Villegas, Z. Sefrioui, M. Varela, E. M. Gonzalez, J. Santamaria, and J. L. Vicent, Phys. Rev. B 69, 134505 (2004).

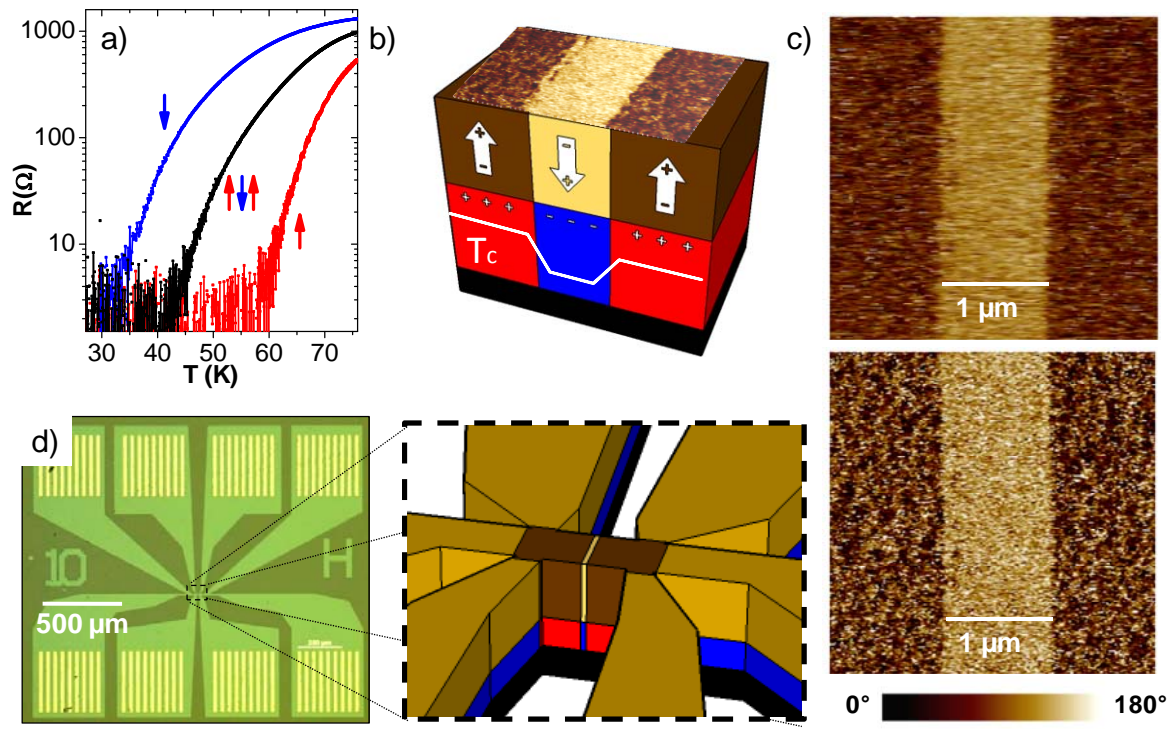

**Figure 1.** (a) Subsequent resistance *vs* temperature measurements of a BFO/ 3 u.c. YBCO bridge fully polarized in the « down » state (blue curve), in the « up » state (red curve), and after the S/S'/S device has been written (black curve). Bias current: 100 nA. (b) Schematic representation of the S/S'/S device. The upper layer represents the BFO with up (brown) and down (yellow) polarized domains. The lower layer represents the accumulated (red) and depleted (blue) YBCO regions. The field-effect induced spatial modulation of the critical temperature $T_C$ is shown. (c) Subsequent PFM phase images ferroelectric domains the day they were written (upper image) and after three weeks (lower image). The color scale shows the PFM phase, in which 0° corresponds to upwards polarization. (d) Microscope picture of the photo-lithographied multi-probe bridge and schematic representation of the area in which the ferroelectric structure is defined to create the S/S'/S junction.

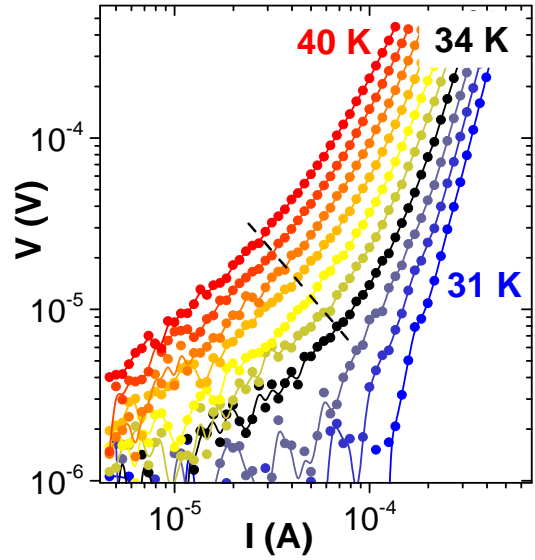

**Figure 2.** Isothermal voltage-current characteristics of a BFO/ 3 u.c. YBCO device patterned with a S/S'/S structure (S' is 1 μm long), for temperatures between 40 K and 31 K spaced by 1 K, under no applied magnetic field.

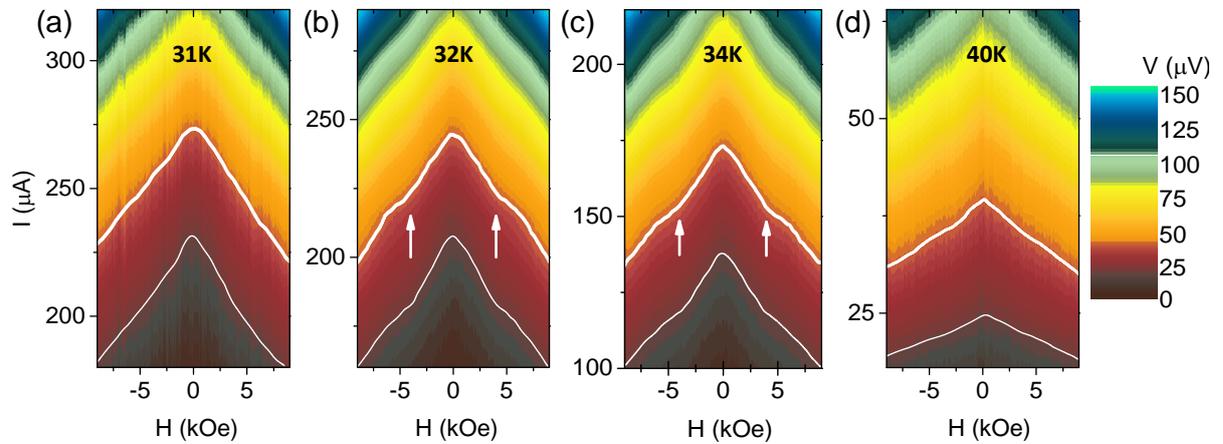

**Figure 3**. Contour plot of the voltage-current characteristics for different applied magnetic fields (ΔH=100 Oe) for a BFO/ 3 u.c. YBCO bridge patterned with a 1 μm junction, at different temperatures (see legends). The color code corresponds to the measured voltage (in V). Interpolated I(H) for V=20 μV and V=40 μV are drawn in white colour as a guide to the eye.

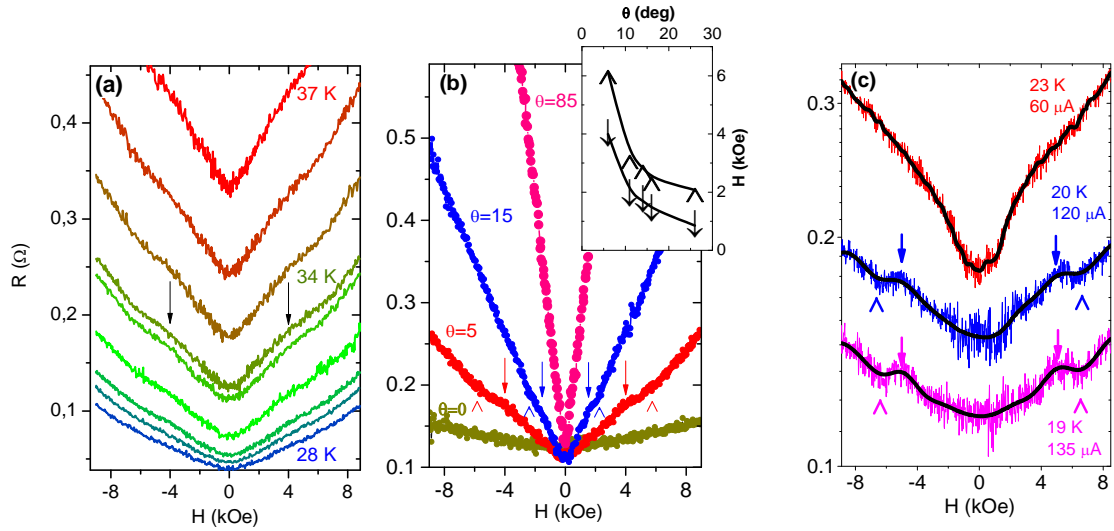

**Figure 4.** Resistance *vs.* applied magnetic field of a BFO/ 3 u.c. YBCO bridge patterned with S/S'/S junction (S' is is 1 µm long). In (a) the field is applied at an angle θ=5°±1° with respect to the YBCO *ab* planes. Temperature ranges between 28 K (blue curve) and 37 K (red curve). The arrows point at the kinks. In (b) the current and temperature are set at 120 µA and 34 K, and the angle θ with respect to the *ab* planes is varied (see legend). Arrows (↓) and arrowheads (∧) point at the inflexion points that delimit the kinks, for θ=5°±1° (red) and θ=15°±1° (blue). Inset: characteristic magnetic fields at which an inflexion points are observed as a function of θ. The lines are a guide to the eye. (c) Resistance *vs.* applied magnetic field for a BFO/ 4 u.c. YBCO bridge in which a S/S'/S structure has been patterned (S' is 1 µm long). H is applied parallel to the YBCO *ab* planes (θ=0°±1°). The superposed black lines are obtained by smoothing (FFT filter) the raw data (shown in color).